\documentclass[11p]{article}

\usepackage{amsfonts}
\usepackage{amsmath}
\usepackage{amscd}
\usepackage{epsf}
\usepackage{times}
\usepackage{latexsym}
\usepackage{graphics}
\usepackage{graphicx}

\begin{document}
\title{Compactons and kink-like solutions of
BBM-like equations \\by means of factorization}
\date{}
\maketitle
\begin{center}
\c{S}. Kuru\\
{\it Department of Physics, Faculty of Sciences, Ankara University,
06100 Ankara, Turkey}
\end{center}
\begin{abstract}
In this work, we study the Benjamin-Bona-Mahony like equations with
a fully nonlinear dispersive term by means of the factorization
technique. In this way we find the travelling wave solutions of this
equation in terms of the Weierstrass function and its degenerated
trigonometric and hyperbolic forms. Then, we obtain the pattern of
periodic, solitary, compacton and kink-like solutions. We give also
the Lagrangian and the Hamiltonian, which are linked to the
factorization, for the nonlinear second order ordinary differential
equations associated to the travelling wave equations.

\end{abstract}

Email:kuru@science.ankara.edu.tr
\section{Introduction}
The nonlinear dispersive equations $K(m,n)$ (or generalized
Korteweg-de Vires equation)
\begin{equation}
u_{t}+(u^m)_x+(u^n)_{xxx}=0, \quad\quad m>0,\,\,1<n\leq 3\label{1.1}
\end{equation}
have been dealt with by Rosenau P. and Hyman J.M. \cite{rosenau}.
They built travelling wave solutions of (\ref{1.1}) with
compact support and called them compactons. Such compactons
are solitary waves but, besides the compact support,
they have some different features from the
solitons. So, while the width of the soliton depends on its amplitude,
the width and amplitude of a compacton are independent.
However, compactons sometimes behave as solitons:
they also collide elastically and their amplitudes
depend on speed of the wave. In the literature there are many
studies on $K(m,n)$ equations and their compacton solutions
\cite{rosenau,rosenau1,wazwaz,wazwaz1,wazwaz2,wazwaz3,taha,ludu}.

The Benjamin-Bona-Mahony (BBM) equation has a higher order
nonlinearity of the form
\begin{equation}
u_{t}+u_{x}+a\,u^m\,u_{x}-u_{xxt}=0, \quad\quad n\geq1\label{1.2}
\end{equation}
where $a$ is constant \cite{benjamin,pilar1}. This equation is an
alternative to the Korteweg-de Vries equation, and describes the
unidirectional propagation of small-amplitude long waves on the
surface of water in a channel, hydromagnetics waves in cold plasma
and acoustic waves in anharmonic crystals. Thus, in this sense, it
has some advantages compared with the Korteweg-de Vries equation.

Now, we will consider the BBM-like equations, $B(m,n)$,  with a
fully nonlinear dispersive term, similar to the $K(m,n)$ equations,
\begin{equation}
u_{t}+u_{x}+a\,(u^m)_x-(u^n)_{xxt}=0, \quad\quad
m,\,n>1\,.\label{1.3}
\end{equation}
There are some studies related with these BBM-like equations
\cite{wazwaz4,wazwaz5,wang,yadong},  in general using the
sine-cosine and the tanh method, in order to get the travelling wave
solutions of this family. In this respect, let us remark that in the
literature there are many methods to obtain exact or special
travelling wave solutions (soliton solutions) of nonlinear equations
\cite{helal}, but there is not a standard method. So, it is
important to investigate new methods to get solutions in different
ways. In this work, we deal with BBM-like equations systematically
by means of the factorization technique
\cite{pilar1,pilar,Perez,pilar2} to find the travelling wave
solutions. When looking for the travelling wave solutions, the
BBM-like equations will reduce to second order nonlinear ordinary
differential equations (ODE). Then, we can apply straightforwardly
the factorization to these ODE's.

\begin{figure}[h]
  \centering
\includegraphics[width=0.4\textwidth]{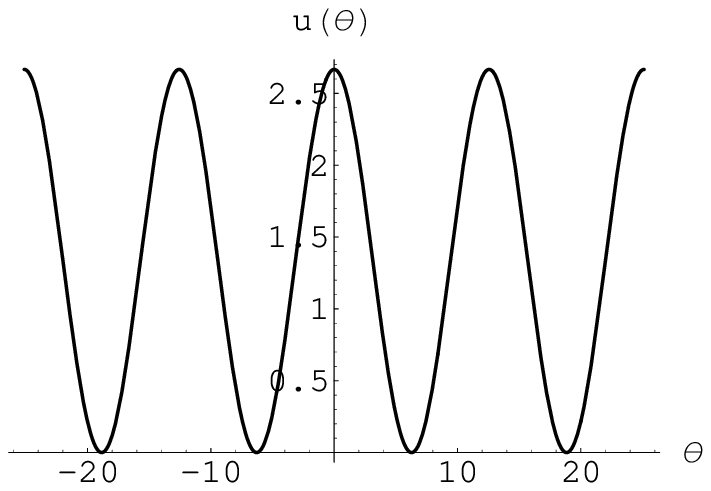}%
\hspace{1cm}%
  \includegraphics[width=0.4\textwidth]{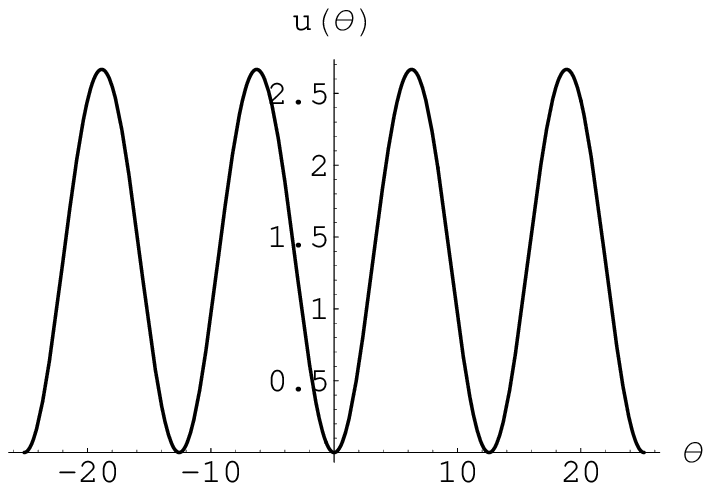}
\caption{Plot of trigonometric solutions (\ref{us}) and (\ref{uc})
for $h=1,\,w=1,\,a=-1,\,n=2$.}
  \label{figuras1}
\end{figure}

\begin{figure}[h]
  \centering
\includegraphics[width=0.4\textwidth]{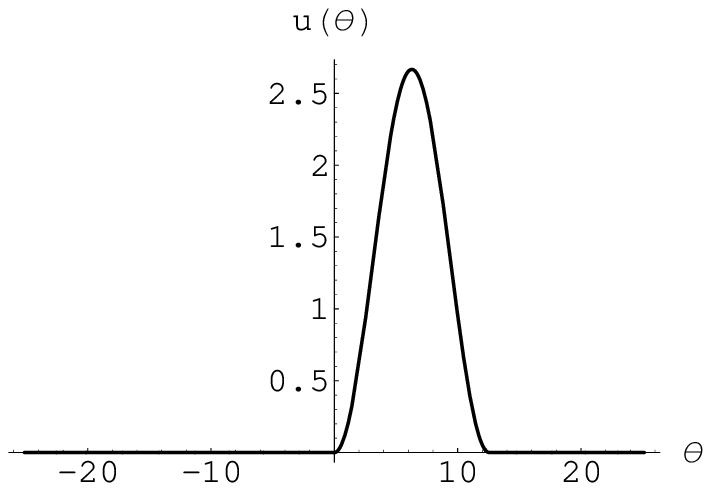}%
\hspace{1cm}%
  \includegraphics[width=0.4\textwidth]{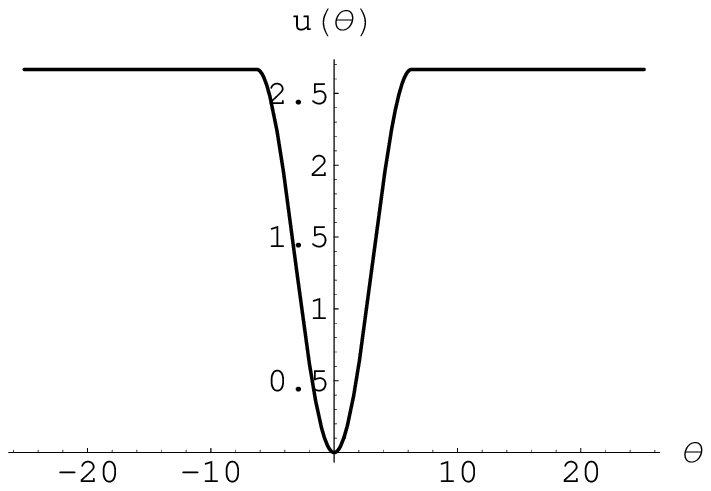}
\caption{Plot of compactons (\ref{42}) and (\ref{44}) for
$h=1,\,w=1,\,a=-1,\,n=2$. The plot of compactons (\ref{43}) and
(\ref{45}) are translations of (\ref{42}) and (\ref{44}),
respectively. }
  \label{figuras2}
\end{figure}

\begin{figure}[h]
  \centering
\includegraphics[width=0.4\textwidth]{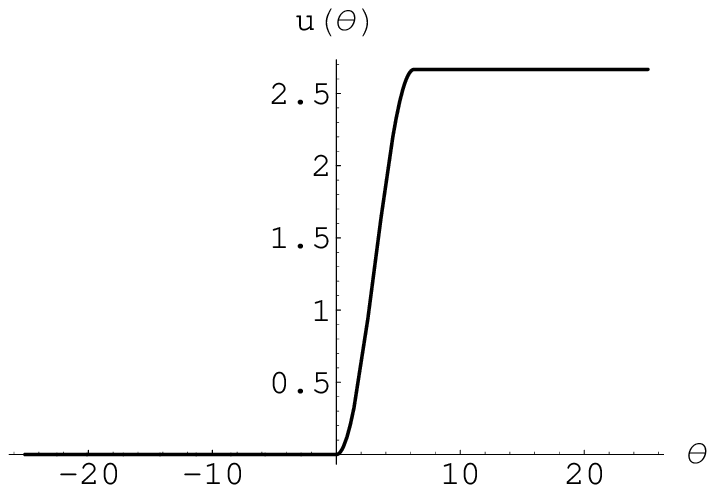}%
\hspace{1cm}%
  \includegraphics[width=0.4\textwidth]{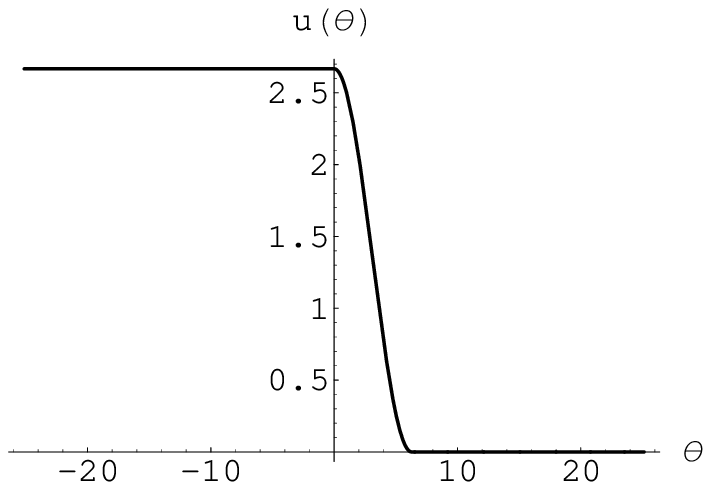}
\caption{Plot of kink-like solutions (\ref{46}) and
(\ref{47}) for $h=1$, $w=1$, $a=-1$, $n=2$.}
  \label{figuras3}
\end{figure}

\begin{figure}[h]
  \centering
\includegraphics[width=0.4\textwidth]{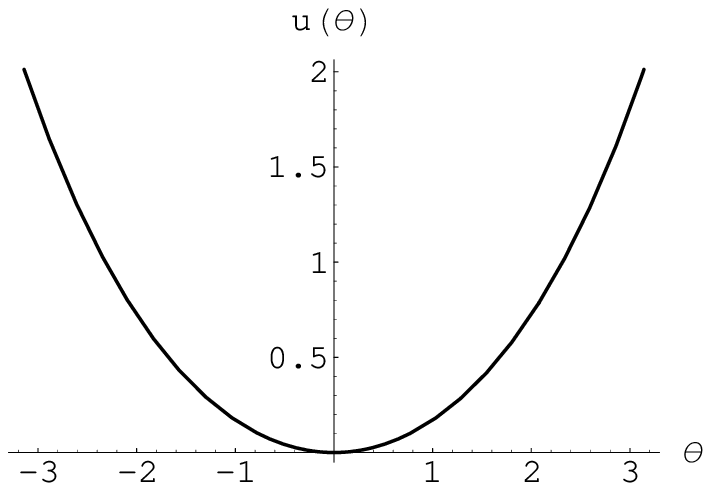}%
\hspace{1cm}%
  \includegraphics[width=0.4\textwidth]{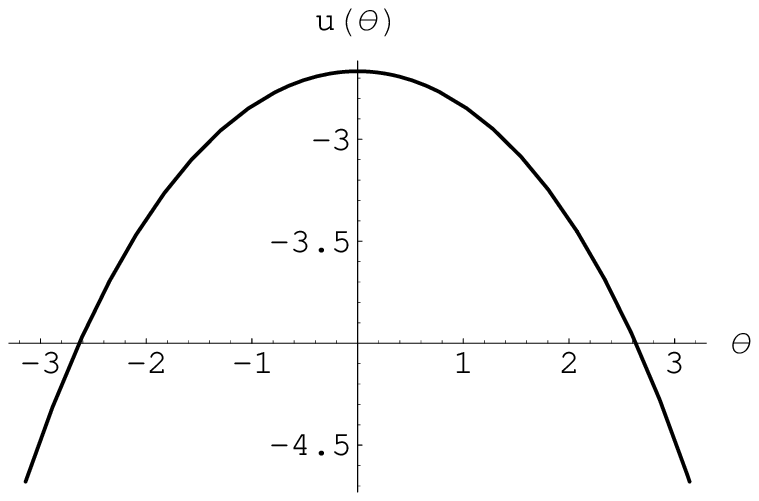}
\caption{Plot of solitary wave pattern solutions (\ref{nes}) and
(\ref{nec}) for $h=1$, $w=1$, $a=1$, $n=2$.}
  \label{figuras4}
\end{figure}

In section 2, we briefly introduce the factorization technique for a
class of nonlinear second order ODE's. Then, in section 3 we apply
this method to the BBM-like equations. In section 4, we obtain the
travelling wave solutions of this equation in terms of Weierstrass
functions which cover the pattern of solitary wave and periodic
solutions. We also construct compactons and kink-like solutions for
this family and we give the graphics of these solutions in
Fig.~\ref{figuras1} (periodic), Fig.~ \ref{figuras2} (compactons),
 Fig.~\ref{figuras3} (kink-like) and Fig.~\ref{figuras4} (solitary
wave pattern). We give the Lagrangian and the Hamiltonian
corresponding to the second order nonlinear ODE and their relation
with the factorization in section 5. Finally, in the last section we
end this work with some conclusions.

\section{Factorization of the nonlinear second
order ODE}

Let us consider, the nonlinear second order ODE of the special form
\begin{equation}\label{9}
 \frac{d^2 W}{d \theta^2}-\beta \frac{d W}{d \theta}+F(W)=0
\end{equation}
where $\beta$ is constant and $F(W)$ is an arbitrary function of
$W$. The factorized form of this equation can be written as
\begin{equation}\label{10}
\left[\frac{d}{d \theta}-f_2(W,\theta)\right]\left[\frac{d}{d
\theta}-f_1(W,\theta)\right] W(\theta)=0\,.
\end{equation}
Here, $f_1$ and $f_2$ are unknown functions that may depend
explicitly on $W$ and $\theta$. Expanding (\ref{10}) and comparing
with (\ref{9}), we obtain the following consistency conditions
\begin{equation}\label{12}
f_1f_2=\frac{F(W)}{W}+\frac{\partial f_1}{\partial \theta}, \qquad
f_2+\frac{\partial(W f_1)}{\partial W}=\beta.
\end{equation}
If we solve (\ref{12}) for $f_{1}$ or $f_{2}$, it will allow us to
write a compatible first order ODE
\begin{equation}\label{14}
\left[\frac{d}{d \theta}-f_1(W,\theta)\right] W(\theta)=0
\end{equation}
that provides a solution for the nonlinear ODE (\ref{9})
\cite{pilar1,pilar,Perez,pilar2}. In the applications of this paper
$f_{1}$ and $f_{2}$ will depend only on $W$.

\section{Factorization of the second order nonlinear ODE's
corresponding to the BBM-like equations
}

Let us assume that (\ref{1.3}) has travelling wave solutions
\begin{equation}\label{15}
u(x,t)=\phi(\xi),\quad\quad \xi=hx+wt
\end{equation}
where $h$ and $w$ are real constants. Substituting (\ref{15}) into
(\ref{1.3}) with the condition $m=n$ and after integrating, we
obtain the second order nonlinear ODE
\begin{equation}\label{16}
(\phi^n)_{\xi\xi}-A\,\phi-B\,\phi^n+D=0
\end{equation}
together with the constants
\begin{equation}\label{17}
A=\frac{h+w}{h^2\,w},\quad\quad B=\frac{a}{h\,w},\quad\quad
D=\frac{R}{h^2\,w}
\end{equation}
here $R$ is integration constant. Now, if we introduce the following
natural transformation of the dependent variable
\begin{equation}\label{18}
\phi^n(\xi)=W(\theta),\quad\quad\xi=\theta
\end{equation}
then, Eq. (\ref{16}) becomes
\begin{equation}\label{19}
\frac{d^2 W}{d \theta^2}-A\,W^{1/n}-B\,W+D=0.
\end{equation}
Next, we will apply the factorization technique to
Eq.~(\ref{19}). In this case, $\beta=0$ and $F(W)$ is
\begin{equation}\label{20}
F(W)=-(A\,W^{1/n}+B\,W-D)
\end{equation}
therefore, the consistency conditions take the form:
\begin{equation}\label{21}
f_1f_2=-A\,W^{(1-n)/n}-B+D\,W^{-1}
\end{equation}
\begin{equation}\label{22}
f_2=-f_1-W\,\frac{df_1}{dW}.
\end{equation}
Substituting (\ref{21}) into (\ref{22}), we get only one
first order consistency equation
\begin{equation}\label{23}
f_1^2+f_1\,W\frac{df_1}{dW}-A\,W^{(1-n)/n}-B+D\,W^{-1}=0
\end{equation}
whose solution gives the unknown function $f_1$,
\begin{equation}\label{24}
f_1(W)=\pm\frac{1}{W}\sqrt{B\,W^2-2\,D\,W+\frac{2\,n\,A}{n+1}\,W^{(n+1)/n}+C}\,
\end{equation}
where $C$ is an integration constant. Then, we can write the first
order ODE (\ref{14})
\begin{equation}\label{25}
\frac{dW}{d
\theta}-\sqrt{B\,W^2-2\,D\,W+\frac{2\,n\,A}{n+1}\,W^{(n+1)/n}+C}=0\,.
\end{equation}
In order to solve this equation for $W$ in a more general way, let
us take $W$ in the form $W=\varphi^p,\,p\neq0,1$ so, the first order
ODE (\ref{25}) can be rewritten in terms of $\varphi$ as
\begin{equation}\label{26}
(\frac{d\varphi}{d
\theta})^2=\frac{B}{p^2}\,\varphi^2-\frac{2\,D}{p^2}\,\varphi^{2-p}+
\frac{2\,n\,A}{(n+1)\,p^2}\,\varphi^{p(\frac{1-n}{n})+2}+\frac{C}{p^2}\,\varphi^{2-2\,p}\,.
\end{equation}
If we want to guarantee the integrability of (\ref{26}), the powers
of $\varphi$ have to be integer numbers between $0$ and $4$
 \cite{ince}. Having
in mind the conditions on $n$ ($n>1$) and $p$ ($p\neq0,1$), we have
the following possible cases:
\begin{itemize}
\item If $C=0,\,\,D=0$, then we can choose $p$ in the following
way
\begin{equation}\label{27}
p\in{\{-\frac{2n}{1-n},-\frac{n}{1-n},\frac{n}{1-n},\frac{2n}{1-n}\}}.
\end{equation}
Here, we deal with only the case $p=-\frac{n}{1-n}$, since the cases
$p=\pm\frac{2n}{1-n}$ and $p=\frac{n}{1-n}$ give rise to the same
solutions for (\ref{1.3}). Thus, Eq.~(\ref{26}) takes the form for
$p=-\frac{n}{1-n}$:
\begin{equation}\label{29}
(\frac{d\varphi}{d \theta})^2=\frac{(n-1)^2\,B}{n^2}\,\varphi^2+
\frac{2\,(n-1)^2\,A}{n\,(n+1)}\,\varphi\,.
\end{equation}
Notice that for the special values of $n=2,3$, we have the $B(2,2)$
and $B(3,3)$ equations and the obtained solutions can be read as
particular cases of this situation.
\item If $C=0$ and $D\neq0$, we have the special cases:
$(p=2,\,n=2)$ and
$(p=-2,\,n=2)$ which correspond to the $B(2,2)$ equations. Since
these two cases also give rise to the same solutions for
(\ref{1.3}), we will consider only one of them. Now,
Eq.~(\ref{26}) has the following form for $(p=2,\,n=2)$
\begin{equation}\label{32}
(\frac{d\varphi}{d \theta})^2=\frac{B}{4}\,\varphi^2+
\frac{A}{3}\,\varphi-\frac{D}{2}\,.
\end{equation}
\end{itemize}

\section{Travelling wave solutions for BBM-like equations}

In this section we will obtain the solutions of the differential
equations (\ref{29}) and (\ref{32}) which allow us  to get the
travelling wave solutions of $B(n,n)$  (\ref{1.3}).
To solve this type of
differential equations,  it will be useful to recall some
properties of the Weierstrass functions  \cite{Bateman,watson}.

Let us consider a differential equation with a quartic polynomial
\begin{equation}\label{ef}
\big(\frac{d\varphi}{d\theta}\big)^2 =P(\varphi) =
a_{0}\,\varphi^4+4\,a_{1}\,\varphi^3+6\,a_{2}\,\varphi^2+4\,a_{3}\,\varphi+a_{4}\,.
\end{equation}
The solution of this equation can be written in terms of the
Weierstrass function $\wp(\theta;g_{2},g_{3})$ where the invariants
$g_2$ and $g_3$ of (\ref{ef}) are
\begin{equation}\label{gg}
g_{2}= a_{0}\,a_{4}-4\,a_{1}\,a_{3}+3\,a_{2}^2,\ \ g_{3}=
a_{0}\,a_{2}\,a_{4}+2\,a_{1}\,a_{2}\,a_{3}-a_{2}^{3}-a_{0}\,a_{3}^2-a_{1}^{2}\,a_{4}.
\end{equation}
Then, the solution $\varphi$ can be found as
\begin{equation}\label{x}
\varphi(\theta)=\varphi_0+\frac{1}{4}P'(\varphi_0)\left(\wp(\theta;g_{2},g_{3})-
\frac{1}{24}P''(\varphi_0)\right)^{-1}
\end{equation}
where the prime ($'$) denotes the derivative with respect to
$\varphi$, and $\varphi_0$ is one of the roots of the polynomial
$P(\varphi)$ (\ref{ef}). The discriminant
$\Delta=g_2^3-27\,g_3^2=0$, allows us to express Weierstrass
functions in terms of trigonometric and hyperbolic functions
\cite{Bateman,watson}
\begin{equation}\label{sh}
\wp(\theta;12\,b^2,-8\,b^3)=b+3\,b \sinh^{-2}[(3\,b)^{1/2}\theta]
\end{equation}
\begin{equation}\label{sn}
\wp(\theta;12\,b^2,8\,b^3)=-b+3\,b \sin^{-2}[(3\,b)^{1/2}\theta].
\end{equation}
Now, we will examine by separate each of the two cases considered in the above section.

\subsection{$C=0,\,D=0$}
In this case we have second order polynomial
\begin{equation}\label{34}
P(\varphi)=\frac{(n-1)^2\,B}{n^2}\,\varphi^2+
\frac{2\,(n-1)^2\,A}{n\,(n+1)}\,\varphi
\end{equation}
whose roots are
\begin{equation}
\varphi_0=0,\quad\quad \varphi_0=-\frac{2\,n\,A}{(n+1)B}\,.
\end{equation}
The invariants have the form
\begin{equation}\label{ggn}
g_{2}= \frac{(n-1)^4\,B^2}{12\,n^4},\ \quad\quad \ g_{3}=
-\frac{(n-1)^6\,B^3}{216\,n^6}
\end{equation}
and discriminant $\Delta=0$.

Then, we can find the solutions from (\ref{x}) in terms of the
Weierstrass functions: for $\varphi_0=0$
\begin{equation}\label{35}
\varphi(\theta)=
-\frac{6\,n\,(n-1)^2\,A}{(n+1)\left[(n-1)^2\,B-12\,n^2\,\wp(\theta;g_2,g_3)\right]}
\end{equation}
 and for $\varphi_0=-\frac{2\,n\,A}{(n+1)B}$
\begin{equation}\label{36}
\varphi(\theta)=
\frac{2\,n\,A}{(n+1)\,B}\left[\frac{2\,(n-1)^2\,B+12\,n^2\,\wp(\theta;g_2,g_3)}
{(n-1)^2\,B-12\,n^2\,\wp(\theta;g_2,g_3)}\right]\,.
\end{equation}

Since the discriminant $\Delta=0$,  we can substitute (\ref{sh})
and (\ref{sn}) into the solutions (\ref{35}) and (\ref{36}),
leading to the following type solutions  of (\ref{29}): a) periodic solutions
\begin{equation}\label{38}
\varphi(\theta)=-\frac{2\,n\,A}{(n+1)\,B}\sin^2
\left({\frac{\sqrt{-B}\,(n-1)}{2\,n}}\theta\right)
\end{equation}
\begin{equation}\label{39}
\varphi(\theta)=-\frac{2\,n\,A}{(n+1)\,B}\cos^2
\left({\frac{\sqrt{-B}\,(n-1)}{2\,n}}\theta\right)
\end{equation}
for $B<0$,  and b) hyperbolic solutions
\begin{equation}\label{40}
\varphi(\theta)=\frac{2\,n\,A}{(n+1)\,B}\sinh^2
\left({\frac{\sqrt{B}\,(n-1)}{2\,n}}\theta\right)
\end{equation}
\begin{equation}\label{41}
\varphi(\theta)=-\frac{2\,n\,A}{(n+1)\,B}\cosh^2
\left({\frac{\sqrt{B}\,(n-1)}{2\,n}}\theta\right)
\end{equation}
for $B>0$. Notice that the first two real solutions (\ref{38}) and
(\ref{39}) are simply related by a transformation.

Now, taking into account (\ref{15}), (\ref{18}) and
$W=\varphi^{p},\,p=n/(n-1)$, the solution of (\ref{1.3}) can be
written as
\begin{equation}\label{37}
u(x,t)=\phi(\xi)=W^{1/n}(\theta)=\varphi^{p/n}(\theta),\quad\quad\theta=\xi=h\,x+w\,t.
\end{equation}
Substituting (\ref{38}) and (\ref{39}) into (\ref{37}), we obtain
periodic solutions of the equation (\ref{1.3})
\begin{equation}\label{us}
u(x,t)=\left[-\frac{2\,n\,A}{(n+1)\,B}\sin^2
\left({\frac{\sqrt{-B}\,(n-1)}{2\,n}}(h\,x+w\,t)\right)\right]^{1/n-1}
\end{equation}
\begin{equation}\label{uc}
u(x,t)=\left[-\frac{2\,n\,A}{(n+1)\,B}\cos^2
\left({\frac{\sqrt{-B}\,(n-1)}{2\,n}}(h\,x+w\,t)\right)\right]^{1/n-1}
\end{equation}
for $B<0$. Then, combining the trivial solution $u(x,t)=0$ with
(\ref{us}) and (\ref{uc}), we have the compact support solutions for
(\ref{1.3}) in the following way \cite{rosenau},
\begin{equation}\label{42}
u(x,t)=\left\{
\begin{array}{ll} \left[-\frac{2\,n\,A}{(n+1)\,B}\sin^2
\left({\frac{\sqrt{-B}\,(n-1)}{2\,n}}(h\,x+w\,t)\right)\right]^{1/n-1},
&
0\leq{\sqrt{-B}}\,(h\,x+w\,t)\leq \frac{2\,n\,\pi}{(n-1)}\\[2.ex]
0, & \rm otherwise\end{array}\right.
\end{equation}
\begin{equation}\label{43}
u(x,t)=\left\{
\begin{array}{ll} \left[-\frac{2\,n\,A}{(n+1)\,B}\cos^2
\left({\frac{\sqrt{-B}\,(n-1)}{2\,n}}(h\,x+w\,t)\right)\right]^{1/n-1},
&
|{\sqrt{-B}}\,(h\,x+w\,t)|\leq \frac{n\,\pi}{(n-1)}\\[2.ex]
0, & \rm otherwise.\end{array}\right.
\end{equation}
For the trivial solution
$u(x,t)=(-\frac{2\,n\,A}{(n+1)\,B})^{1/n-1}$, we also get the
compactons:
\begin{equation}\label{44}
u(x,t)=\left\{
\begin{array}{ll} \left[-\frac{2\,n\,A}{(n+1)\,B}\sin^2
\left({\frac{\sqrt{-B}\,(n-1)}{2\,n}}(h\,x+w\,t)\right)\right]^{1/n-1},
&
|{\sqrt{-B}}\,(h\,x+w\,t)|\leq \frac{n\,\pi}{(n-1)}\\[2.ex]
(-\frac{2\,n\,A}{(n+1)\,B})^{1/n-1}, & \rm
otherwise\end{array}\right.
\end{equation}
\begin{equation}\label{45}
u(x,t)=\left\{
\begin{array}{ll} \left[-\frac{2\,n\,A}{(n+1)\,B}\cos^2
\left({\frac{\sqrt{-B}\,(n-1)}{2\,n}}(h\,x+w\,t)\right)\right]^{1/n-1},
&
0\leq{\sqrt{-B}}\,(h\,x+w\,t)\leq \frac{2\,n\,\pi}{(n-1)}\\[2.ex]
(-\frac{2\,n\,A}{(n+1)\,B})^{1/n-1}, & \rm
otherwise.\end{array}\right.
\end{equation}

Now, we can construct kink-like solutions
\cite{dusuel}, combining the non trivial solutions (\ref{us}) and
(\ref{uc}) with both (different) constant solutions, $u(x,t)=0$ and
$u(x,t)=(-\frac{2\,n\,A}{(n+1)\,B})^{1/n-1}$, as follows
\begin{equation}\label{46}
u(x,t)=\left\{
\begin{array}{ll}
0,&{\sqrt{-B}}\,(h\,x+w\,t)<0\\[2.ex]
\left[-\frac{2\,n\,A}{(n+1)\,B}\sin^2
\left({\frac{\sqrt{-B}\,(n-1)}{2\,n}}(h\,x+w\,t)\right)\right]^{1/n-1},
&
0\leq{\sqrt{-B}}\,(h\,x+w\,t)\leq \frac{n\,\pi}{(n-1)}\\[2.ex]
(-\frac{2\,n\,A}{(n+1)\,B})^{1/n-1}, & {\sqrt{-B}}\,(h\,x+w\,t)>
\frac{n\,\pi}{(n-1)}\end{array}\right.
\end{equation}
\begin{equation}\label{47}
u(x,t)=\left\{
\begin{array}{ll}
(-\frac{2\,n\,A}{(n+1)\,B})^{1/n-1},&{\sqrt{-B}}\,(h\,x+w\,t)<0\\[2.ex]
\left[-\frac{2\,n\,A}{(n+1)\,B}\cos^2
\left({\frac{\sqrt{-B}\,(n-1)}{2\,n}}(h\,x+w\,t)\right)\right]^{1/n-1},
&
0\leq{\sqrt{-B}}\,(h\,x+w\,t)\leq \frac{n\,\pi}{(n-1)}\\[2.ex]
0, & {\sqrt{-B}}\,(h\,x+w\,t)>
\frac{n\,\pi}{(n-1)}\,.\end{array}\right.
\end{equation}

When we put (\ref{40}) and (\ref{41}) in (\ref{37}), we have the
solitary wave pattern solutions of hyperbolic type
for (\ref{1.3}). In order to get
real solutions, it is necessary to examine the solutions for $n$
even and $n$ odd. Then, for $n$ even and $B>0$ we have the solutions
\begin{equation}\label{nes}
u(x,t)=\left[\frac{2\,n\,A}{(n+1)\,B}\sinh^2
\left({\frac{\sqrt{B}\,(n-1)}{2\,n}}(h\,x+w\,t)\right)\right]^{1/n-1}
\end{equation}
\begin{equation}\label{nec}
u(x,t)=\left[-\frac{2\,n\,A}{(n+1)\,B}\cosh^2
\left({\frac{\sqrt{B}\,(n-1)}{2\,n}}(h\,x+w\,t)\right)\right]^{1/n-1}\,.
\end{equation}
However, if $n$ is odd, we have the solution (\ref{nes}) provided
$A>0$, while the solution (\ref{nec}) is valid only when $A<0$.

The solutions for the  $B(2,2)$ and the  $B(3,3)$ can be easily read
from the above solutions of (\ref{1.3}).

\subsection{$C=0,\,D\neq0$}
The case $n=2$ corresponds to the $B(2,2)$ equation and, according to
(\ref{32}) we have another kind of solutions for this equation.
In this case, the polynomial
\begin{equation}\label{48}
P(\varphi)=\frac{B}{4}\,\varphi^2+
\frac{A}{3}\,\varphi-\frac{D}{2}\,
\end{equation}
has two roots
\begin{equation}\label{49}
\varphi_0^{\pm}=\frac{-A\pm\sqrt{4\,A^2+18\,B\,D}}{3\,B}\,.
\end{equation}
From (\ref{gg}), the invariants for Weierstrass functions take the
values
\begin{equation}\label{gg2}
g_{2}= \frac{B^2}{192},\ \quad\quad \ g_{3}= -\frac{B^3}{13824}
\end{equation}
and the discriminant also vanishes $\Delta=0$.
Then, from (\ref{x}) the solutions read
\begin{equation}\label{50}
\varphi^{\pm}=-\frac{2\,A}{3\,B}\mp\frac{\sqrt{4\,A^2+18\,B\,D_2}}{3\,B}
\left(\frac{5\,B+48\,\wp(\theta;g_2,g_3)}{B-48\,\wp(\theta;g_2,g_3)}\right)\,.
\end{equation}
Since the discriminant equals zero, we can express these solutions
in terms of trigonometric and hyperbolic functions using the
relations (\ref{sh}) and (\ref{sn}):
\begin{equation}\label{51}
\varphi^{\pm}=-\frac{2\,A}{3\,B}\pm\frac{\sqrt{4\,A^2+18\,B\,D}}{3\,B}\,
\cos(\frac{\sqrt{-B}}{2}\,\theta)
\end{equation}
for $B<0$ and
\begin{equation}\label{52}
\varphi^{\pm}=-\frac{2\,A}{3\,B}\pm\frac{\sqrt{4\,A^2+18\,B\,D}}{3\,B}\,
\cosh(\frac{\sqrt{B}}{2}\,\theta)
\end{equation}
for $B>0$. Hence, from (\ref{37}) we have
$u(x,t)=\phi(\xi)=\varphi(\theta),\,\theta=\xi=h\,x+w\,t$, and the
solitary wave pattern and trigonometric solutions of (\ref{1.3}) can
be found substituting (\ref{51}) and (\ref{52}) into this
expression. Here, again combining the trivial solution
$u(x,t)=-\frac{2\,A}{3\,B}$ with the trigonometric solution, we can
construct compactons as
\begin{equation}\label{53}
u^{\pm}(x,t)=\left\{
\begin{array}{ll} \left[-\frac{2\,A}{3\,B}\pm\frac{\sqrt{4\,A^2+18\,B\,D}}{3\,B}\,
\cos(\frac{\sqrt{-B}}{2}\,(h\,x+w\,t))\right], &
|{\sqrt{-B}}\,(h\,x+w\,t)|\leq \pi\\[2.ex]
-\frac{2\,A}{3\,B}, & \rm otherwise\, .\end{array}\right.
\end{equation}
Kink-like solutions can also be obtained by taking into account
another trivial solution
$u(x,t)=-\frac{2\,A}{3\,B}\pm\frac{\sqrt{4\,A^2+18\,B\,D}}{3\,B}$
together with $u(x,t)=-\frac{2\,A}{3\,B}$:
\begin{equation}\label{54}
u^{\pm}(x,t)=\left\{
\begin{array}{ll}
-\frac{2\,A}{3\,B},&{\sqrt{-B}}\,(h\,x+w\,t)<-\pi\\[2.ex]
\left[-\frac{2\,A}{3\,B}\pm\frac{\sqrt{4\,A^2+18\,B\,D}}{3\,B}\,
\cos(\frac{\sqrt{-B}}{2}\,(h\,x+w\,t))\right], &
-\pi\leq{\sqrt{-B}}\,(h\,x+w\,t)\leq 0\\[2.ex]
-\frac{2\,A}{3\,B}\pm\frac{\sqrt{4\,A^2+18\,B\,D}}{3\,B}, &
{\sqrt{-B}}\,(h\,x+w\,t)>0\end{array}\right.
\end{equation}
where $4\,A^2+18\,B\,D > 0$ and $B<0$. The solitary wave (hyperbolic
type) pattern solutions are
\begin{equation}\label{55}
u^{\pm}(x,t)=-\frac{2\,A}{3\,B}\pm\frac{\sqrt{4\,A^2+18\,B\,D}}{3\,B}\,
\cosh(\frac{\sqrt{B}}{2}\,(h\,x+w\,t))
\end{equation}
with the conditions  $4\,A^2+18\,B\,D > 0$ and $B>0$.

\section{Lagrangian and Hamiltonian}
From the second order nonlinear ODE (\ref{19}), the Lagrangian can
be written as
\begin{equation}\label{lag}
L_W(W,W_{\theta},\theta)=\frac{1}{2}\left(W_\theta^2+\frac{2\,n\,A}{n+1}\,W^{\frac{n+1}{n}}+B\,W^2-
2\,D\,W\right)\,.
\end{equation}
Then, the canonical momentum is $P_W=\frac{\partial L_W}{\partial
W_\theta}=W_\theta\label{p} $, and the Hamiltonian $H_W=W_\theta P_W
-L_W$ has the form
\begin{equation}H_W(W,P_W,\theta)=\frac{1}{2}\left(P_W^2-\frac{2\,n\,A}{n+1}\,W^{\frac{n+1}{n}}-
B\,W^2+2\,D\,W\right) .\label{hh}\end{equation}  Since $H_W$ does
not depend on the variable $\theta$, it is a constant of motion
$H_W=E$
\begin{equation}E=\frac{1}{2}\left(\left(\frac{dW}{d\,\theta}\right)^2-
\frac{2\,n\,A}{n+1}\,W^{\frac{n+1}{n}}-B\,W^2+2\,D\,W\right)\,.\label{ee}\end{equation}
This equation also gives rise to the first order ODE (\ref{25}) with
the identification $C=2\,E$. We can express this constant of motion
$H_W=E$ as a product of two independent constants of motion
\begin{equation}E=\frac{1}{2}I_1\,I_2\label{e}\end{equation}
where
\begin{equation}I_1=(W_\theta-\sqrt{\frac{2\,n\,A}{n+1}\,W^{\frac{n+1}{n}}+
B\,W^2-2\,D\,W})\,e^{S(\theta)}\label{ý1}\end{equation}
\begin{equation}I_2=(W_\theta+\sqrt{\frac{2\,n\,A}{n+1}\,W^{\frac{n+1}{n}}+
B\,W^2-2\,D\,W})\,e^{-S(\theta)}\label{ý2}\end{equation} and
$S(\theta)$ has the form such that $I_1$ and $I_2$ satisfy
$dI_j/d\theta=0,\,j=1,2$
$$ S(\theta)=\int \frac{A\,W^{\frac{1}{n}}+B\,W-D}{\sqrt{\frac{2\,n\,A}{n+1}\,W^{\frac{n+1}{n}}+
B\,W^2-2\,D\,W}}\,d\theta.$$

\section{Conclusions}
In this work, we have investigated the travelling wave solutions of
the BMM-like equations by means of the factorization technique. We
have factorized the nonlinear second order ODE's corresponding to
the BMM-like equations. Then, we have obtained the solutions in
terms of Weierstrass functions giving rise trigonometric (periodic)
and hyperbolic type solutions (solitary wave pattern) and we have
constructed compactons and kink-like solutions for the BBM-like
equations. In addition to these we give the Lagrangian and the
Hamiltonian corresponding to the second order ODE. We have seen that
the first order ODE which gives rise to the solutions, can also be
obtained from the constant of motion $E$ corresponding to the
Hamiltonian. Then, we have written this constant of motion $E$  as a
product of two independent constants of motion. We note that this
technique is more systematic than others previously used for the
analysis of this type equations. In general, we have more general
solutions and we have recovered all the solutions previously
reported \cite{wazwaz5,wang,yadong}. Finally, we must mention that
this method can be easily implemented to the other nonlinear
equations, in particular we plan to give some results on the
$B(m,n)$ equations with $m\neq n$ in a future publication.

 \section*{Acknowledgments}
The author acknowledges to Dr. Javier Negro for useful discussions.

\end{document}